\documentclass[a4paper]{article}

\usepackage{icrc2013}
\usepackage{amssymb,amsmath}
\usepackage{lscape}
\usepackage{longtable}
\usepackage{ctable}
\usepackage{multirow}
\usepackage{siunitx}
\usepackage[english]{babel}

\title{Spatial Development of Energetic Particle Spectra in Pulsar Wind Nebulae}

\shorttitle{Evolution of Spectra in PWNe}

\authors{
Michael Vorster
and Harm Moraal
}

\afiliations{
Centre for Space Research, North-West University, Potchefstroom Campus, 2520 Potchefstroom, South Africa \\
}

\email{12792322@puk.ac.za}

\abstract{
The evolution of the non-thermal emission emitted by pulsar wind nebulae is generally calculated using spatially-independent particle transport models. Although useful, these models implicitly assume that the source of non-thermal particles is located uniformly throughout the system, contrary to the nature of pulsar wind nebulae where the source is located at the centre of the system. Additionally, it is not possible to take into account the spatial properties of the magnetic field and flow velocity, or the effects of diffusion and gradient and curvature drifts in these models. In this paper we present an extension to the current nebula models by including a spatial dependence in our numerical solutions of a Fokker-Planck particle transport equation. These solutions also show the effect that the magnetic field structure has on the evolution of the particle spectrum. Although applied within the context of pulsar wind nebulae, the presented solutions are equally valid for any similar central source system such as globular clusters.}


\begin{document}
\maketitle

\section{Introduction}

Relativistic particles transported through a system are generally subjected to a number of processes, leading to a spatial and temporal variation in the particle energy spectrum.  A natural way to describe the evolution of the energy spectrum is to use a Fokker-Planck transport equation.  Notable examples include the transport equation used to describe the propagation of cosmic rays in the Galaxy \cite{Ginzburg1964}, and the Parker transport equation used to describe the modulation of cosmic rays in the heliosphere \cite{Parker1965}.

This paper presents the evolution of a non-thermal particle energy spectrum originating form a central source.  The prototypical example of such a system is a pulsar wind nebulae (PWN), with the Crab Nebula being the best-known example.  

It is generally believed that pulsars produce highly relativistic winds that consist of electrons and positrons \cite{Kirk2009}, and possibly hadrons \cite{Cheng1980}.  Due to the plasma nature of the wind, the ideal magnetohydrodynamic limit is satisfied, leading to the pulsar's magnetic field being frozen into the out-flowing wind \cite{Kirk2009}.  It has been shown that a magnetic field frozen into a plasma wind that originates from a rotating magnetic dipole will have an Archimedean spiral structure (e.g., \cite{Parker1958,Michel1973}), and as a result of the large angular velocity of pulsars, this field can be approximated as purely azimuthal.  The exception is at the poles, where the magnetic field has only a radial component.  

The dipolar magnetic field of the pulsar further implies that the direction of the frozen-in magnetic field must reverse between the northern and southern hemispheres of the system.  The consequence of this is that a magnetically-neutral region should exist that separates the regions of opposing magnetic polarity \cite{Coroniti1990}.  If the magnetic and rotation axes of the pulsar are aligned, this region forms a flat sheet in the equatorial plane of the system.  On the other hand, this neutral sheet will have a waved structure if these axes are not aligned.  

When the ram pressure of the pulsar wind is equal to the confining pressure of the ambient medium, a termination shock is formed \cite{Rees1974} where the charged particles are accelerated \cite{Kennel1984}.  Downstream of the termination shock the leptons (electrons and positrons) interact with the frozen-in magnetic field, leading to synchrotron radiation that is observed from radio to X-ray wavelengths.  Additionally, the leptons can also inverse Compton (IC) scatter ambient photons to high-energy and very-high-energy gamma-ray wavelengths \cite{Dejager2009}.  These ambient photons can have a number of origins, including the cosmic microwave background radiation, infra-red radiation from dust, starlight, and even the radiated synchrotron photons.  This non-thermal emission leads to a luminous nebula, or PWN. 

A second example of a central source system is globular star clusters.  It is believed that non-thermal particles are injected into the cluster by millisecond pulsars located at its centre.  The large size of the cluster compared to the compact core of pulsars makes it possible to approximate the central pulsars as a single source.  The particles that diffuse away from the centre of the cluster will produce synchrotron radiation, as well as IC emission \cite{Venter2009}. 
  
The simulations presented in this paper will focus on two key aspects.  The first is the effect that diffusion has on the spatial evolution of a particle spectrum in a PWN.  For these simulations, a spherically-symmetric steady-state transport equation is solved numerically with the spatial transport processes of convection and diffusion taken into account, along with the the energy loss processes of adiabatic cooling, synchrotron radiation, and IC scattering.  

For the second part of the simulations, the effect of drift on the evolution of the spectra is investigated.  Drift processes are related to the geometry of the magnetic field in the system, and to include these processes requires an additional spatial dimension.  The transport equation is thus solved in an axisymmetric steady-state system where particles are transported by convection and diffusion, as well as gradient and curvature drift.  The simulations also take into account the fact that particles will drift along the neutral sheet.

\section{The axisymmetric transport equation}

Let $p$ denote momentum and $\mathbf{r}=\mathbf{r}\left(r,\theta\right)$ the spatial position of a particle.  Synchrotron and IC losses are described by 
\begin{equation}\label{eq:Thom}
\frac{\langle\dot{p}\rangle_{\rm{n-t}}}{p} = -z(\mathbf{r},t)p,
\end{equation}
where
\begin{equation}\label{eq:a_coef}
z(\mathbf{r},t) = \frac{4\sigma_{\rm{T}}}{3\left(m_0c\right)^2}\left(U_B+U_{\rm{IC}} \right)
\end{equation}
is a function of the magnetic $U_B$ and photon $U_{\rm{IC}}$ energy field densities.  The other constants in (\ref{eq:a_coef}) are the Thomson scattering cross-section $\sigma_{\rm{T}}$, the rest mass of the particle $m_0$, and the speed of light $c$.  It follows from (\ref{eq:Thom}) that synchrotron radiation and IC scattering effect the same evolution on the transport equation, and the term "synchrotron losses/radiation" will therefore be used as a collective term that refers to both of these non-thermal energy loss processes.  

It is generally assumed in PWN models (e.g., \cite{Rees1974, Kennel1984, Vanderswaluw2001}) that the convection velocity of the wind has only a radial component $V_r$, with the present simulations following suit.  In the axisymmetric model adiabatic losses are therefore given by 
\begin{equation}
\frac{\langle\dot{p}\rangle_{\rm{ad}}}{p} = -\frac{1}{3r^2}\left(r^2V_r\right). 
\end{equation}  

While it is possible to formally include gradient and curvature drifts in the transport equation, \cite{Parker1965} and \cite{Axford1965} have shown that it is also possible to include these processes directly into the diffusion tensor $\mathbf{\underline{K}}$.  This tensor contains only four elements in an axisymmetric system, with the diagonal elements ($\kappa_{rr}$ and $\kappa_{\theta\theta}$) describing diffusion, and the off-diagonal elements ($\kappa_{r\theta}$ and $\kappa_{\theta r}$) describing gradient and curvature drifts.    

For the simulations it is assumed that the spatial and momentum dependence of the diffusion coefficients $\kappa_{ii}$ can be separated, and that these coefficients scale linearly with momentum, i.e.,
\begin{equation}
\kappa \equiv \kappa_0\kappa\left(r,\theta\right)\beta p,
\end{equation}
where $\beta\approx 1$ is the usual relativistic factor, and $\kappa_0$ a normalisation constant.  Furthermore, the spatial dependence is determined by the relation $\kappa\left(r,\theta\right) \propto 1/B$, where $B$ is the magnetic field.  

It follows from the Fokker-Planck transport equation that the spatial evolution of a particle spectrum in an axisymmetric system can be described by 
\begin{equation}\label{eq:te}
\begin{split}
&\kappa_{rr}\dfrac{\partial^2 f}{\partial r^2} + \dfrac{\kappa_{\theta\theta}}{r^2}\dfrac{\partial^2 f}{\partial \theta^2}\\
&+\left[\dfrac{1}{r^2}\dfrac{\partial}{\partial r}\left(r^2\kappa_{rr}\right) + \dfrac{1}{r\sin\theta}\dfrac{\partial}{\partial \theta}\left(\sin\theta\kappa_{\theta r}\right) -V_r\right]\dfrac{\partial f}{\partial r}\\
&+\left[\dfrac{1}{r^2}\dfrac{\partial}{\partial r}\left(r\kappa_{r\theta}\right) + \dfrac{1}{r^2\sin\theta}\dfrac{\partial}{\partial \theta}\left(\sin\theta\kappa_{\theta\theta}\right)\right]\dfrac{\partial f}{\partial\theta}\\
&+\left[\frac{1}{3r^2}\frac{\partial}{\partial r}\left(r^2V_r\right) +z p\right]\frac{\partial f}{\partial \ln p}+4zpf \\
& =0.
\end{split}
\end{equation}
Here $f(r,p,t)$ denotes the omni-directional distribution function that is related to the more familiar particle density through $N(\mathbf{r},p,t) = 4\pi p^2 f(\mathbf{r},p,t)$.      

Based on X-ray observations (e.g., \cite{Gaensler1999, Mangano2005, Schock2010}), the source spectrum injected into the system is chosen to be $f \propto p^{-4}$.  As the particles under consideration are relativistic ($E\approx pc$), the source spectrum can equivalently be expressed as $N \propto E^{-2}$.

\section{The effect of diffusion on the evolution of the spectra}

The results presented in this section focus specifically on the effect that diffusion has on the evolution of the particle spectra in a PWN where convection, diffusion, adiabatic losses, and synchrotron radiation are important.  For a discussion on the effect of the individual processes on spectral evolution, \cite{Vorster2013}, and the references therein, can be consulted.  Note that the results presented in this section have also been published in \cite{Vorster2013}.  Here more details are also given on the \emph{Crank-Nicolson} numerical scheme that is used to solve the spherically-symmetric version of (\ref{eq:te}).

The velocity and magnetic profiles used in the simulations are related through the ideal magnetohydrodynamic limit $\nabla\times\mathbf{V}\times\mathbf{B}=0$.  For a radial flow and purely azimuthal magnetic field, this leads to 
\begin{equation}\label{eq:MHD_limit_reduce}
VBr=V_0B_0r_0,
\end{equation}
where $V_0, B_0$, and $\kappa_0$ represent the values at the termination shock $r_0$.  For the simulations the profile $V\propto r^{-0.5}$ is used, and it therefore follows from (\ref{eq:MHD_limit_reduce}) that $B\propto r^{-0.5}$.  In a spherically-symmetric system the diffusion tensor reduces to a single isotropic diffusion coefficient $\kappa$.  From the scaling $\kappa \propto 1/B$ it follows that $\kappa \propto r^{0.5}$.  The values used for the variables can be found in \cite{Vorster2013}.  

 \begin{figure}[!t]
  \centering
  \includegraphics[scale=0.7]{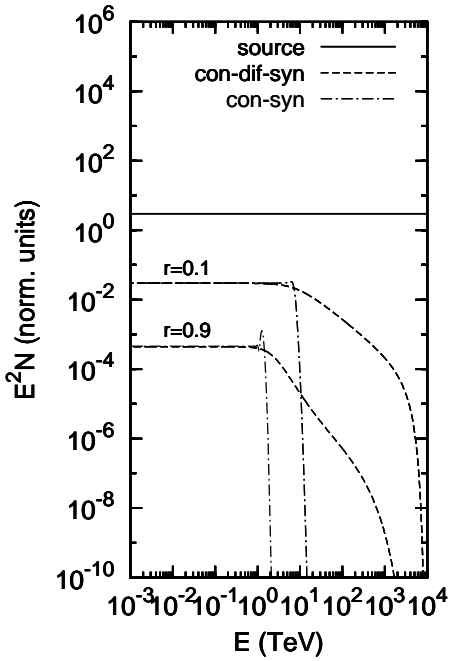}
  \caption{The evolution of the spectra in a spherically-symmetric system.  The particles are transported by convection and diffusion, while being subjected to adiabatic cooling, synchrotron radiation, and IC scattering (con-dif-syn).  Shown for comparison are the spectra for a system where diffusion is absent (con-syn).  The spectra are shown at scaled radial distances, with $r$ scaled to the size of the system.  The outer boundary of the system is therefore located at $r=1$.}
  \label{fig:conv-diff-sync}
 \end{figure}

\vspace{0.4cm}
\subsection{Model predictions}

Figure \ref{fig:conv-diff-sync} shows the evolution of the spectra in the PWN, as predicted by the model.  Note that the spectra are plotted at normalised distances, where $r$ is normalised with respect to the size of the PWN.

As $\kappa \propto \beta p$, the particles that suffer the most synchrotron losses will also be the particles that are most affected by diffusion.  Compared to a system where only convection is present, diffusion will reduce the propagation time through the system, thereby decreasing the synchrotron losses suffered by the leptons.  The result is that the characteristic synchrotron cut-off that should appear at higher energies is transformed into a much harder spectrum.  It is only at the the highest energies ($E \gg 1\,\text{TeV}$) that this cut-off appears.  At lower energies the particles are primarily transported by convection, and the particles subsequently only suffer adiabatic cooling.  The result is a spectrum that decreases in intensity, while the spectral index remains the same as that of the source spectrum.    

In the outer part of the system ($r=0.9$) the spectrum in the energy range $1 \mbox{ TeV}\lesssim E \lesssim 10$ TeV initially softens, and can be fitted with a power-law $N\propto E^{-3.8}$.  This initial softening of the spectrum is followed by a marginally harder spectrum at $E \sim 10$ TeV, and the spectrum in the energy range $10 \mbox{ TeV}\lesssim E \lesssim 200$ TeV can again be fitted with a power law $N \propto E^{-3.5}$.  At larger energies ($E \gtrsim 100$ TeV) the spectrum softens again,  signifying the beginning of the high-energy synchrotron cut-off.

\subsection{Comparison with observations}

Observations focussing on the inner region of Vela X show a bright X-ray nebula \cite{Mangano2005}.  The synchrotron photon index $\Gamma$ in the $3-10$ keV range was extracted from a number of annular regions of increasing size, revealing a softening of the index with increasing distance from the shock ($r_0=0.35'$ \cite{Ng2004}.)  In the inner region ($r \le 0.5'$) the photon index was found to be $\Gamma=1.50\pm 0.02$, while the index in the outer annular region ($8'\le r\le 12'$) was found to be $\Gamma=1.90\pm 0.06$.  If the energy spectrum of the particles is described by $N \propto E^{-\alpha}$, then the relation $\Gamma=(\alpha+1)/2$ implies $\alpha=2.00\pm 0.04$ in the inner annular region, and $\alpha=2.80\pm 0.12$ in the outer region.  

A similar X-ray observation ($0.5-9$ keV) has also been performed for the nebula MSH 15-52 \cite{Schock2010}.  In the inner annular region ($30''\le r \le 57''$) the photon index was found to be $\Gamma=1.66\pm 0.02 \hspace{0.2cm}(\alpha=2.32\pm 0.04)$, softening to a value of $\Gamma=2.24\pm 0.28 \hspace{0.2cm}(\alpha=3.48\pm 0.56)$ in the outer annular region ( $246'' \le r \le 300''$).       

The electron energy needed to produce a synchrotron photon with an energy of $E_{\rm{keV}}$ is \cite{Dejager2009}
\begin{equation}\label{eq:min_energy}
E \approx (220\mbox{ TeV})B_{\mu\rm{G}}^{-1/2}E_{\rm{keV}}^{1/2}.
\end{equation}
Using the same parameters as those chosen by \cite{Vorster2013}, the magnetic field has the values $B_{r=0.1}=113$ $\mu$G and $B_{r=0.9}=38$ $\mu$G.  From (\ref{eq:min_energy}) it follows that the electron energy needed to produce synchrotron emission in the energy range $E_{\rm{keV}}=0.5-10$ keV is $E_{r=0.1}=15-65$ TeV and $E_{r=0.9}=25-113$ TeV.  Figure \ref{fig:conv-diff-sync} shows that these are the energy ranges where diffusion will have the largest influence on the evolution of the spectra.  Furthermore, the index $\alpha_{r=0.9}=3.5$ derived from Figure \ref{fig:conv-diff-sync} (in the energy range $2 \mbox{ TeV} \lesssim E \lesssim 100$ TeV) is consistent with the observed index in the outer regions of MSH 15-52.  Selecting the wind profile $V \propto 1/r^{0.1}$, while keeping all the other parameters fixed, the model calculates a particle index of $\alpha_{r=0.9}=2.8$, in agreement with the observed index in the outer regions of the compact Vela PWN.

\section{The effect of drift on spectral evolution}

For these simulations the axisymmetric transport equation (\ref{eq:te}) includes the effects of convection, diffusion, adiabatic losses, and drift.  The last process includes gradient, curvature, and neutral sheet drift.  As mentioned previously, the first two processes are included into the diffusion tensor, while neutral sheet drift is included as explained in \cite{Burger1987}.  For the purposes of illustration, a flat neutral sheet is used in the simulations.  The transport equation is solved numerically using the \emph{Douglas} Alternating Direction Implicit scheme \cite{Douglas1962}.  

It should be noted that drift is dependent on the product $qA$, where $q$ is the electric charge of the particle, and 
\begin{equation}
A \equiv \cos\alpha = \frac{\pmb{\Omega}\cdot\pmb{\mu}}{\Omega\mu}.
\end{equation}
If $qA>0$, particles will drift from the polar to equatorial region and out along the neutral sheet, and vice versa for $qA<0$.  For a graphical illustration of the expected drift motion, \cite{Pesses1981} can be consulted.

 \begin{figure}[!t]
  \centering
  \includegraphics[scale=0.4]{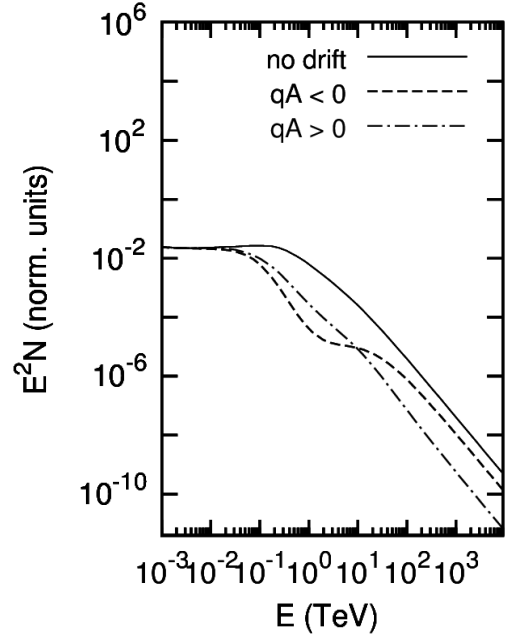}
  \caption{The calculated spectra at $r=0.9$ and $\theta=45^{\circ}$ for a system where convection, diffusion, adiabatic losses, and gradient, curvature, and neutral sheet drift are present.  Shown for comparison is the spectrum for a system where drift is absent.}
  \label{fig:t_45}
 \end{figure}

 \begin{figure}[!t]
  \centering
  \includegraphics[scale=0.4]{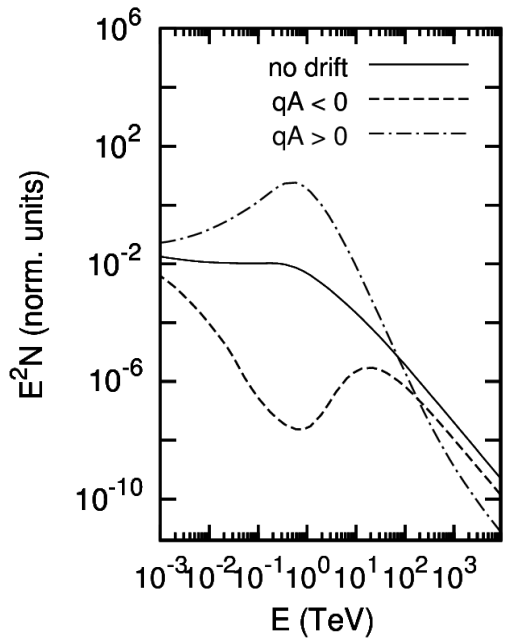}
  \caption{The same as Figure \ref{fig:t_45}, but with the spectra plotted at $\theta=90^{\circ}$.}
  \label{fig:t_90}
 \end{figure}

Figure \ref{fig:t_45} shows the spectra at $r=0.9$ and the angular position of $\theta=45^{\circ}$.  Shown for comparison in the same figure is the spectrum calculated for a scenario where drift is absent.  In the $qA>0$ scenario drift reduces the intensity when $E\gtrsim 10\,\text{TeV}$, but does not change the spectral shape.  In the energy range $0.1\,\text{TeV} \lesssim E \lesssim 10\,\text{TeV}$ the spectrum is marginally harder.  For the $qA<0$ scenario the spectrum at $E\gtrsim 10\,\text{TeV}$ is similar to the no-drift spectrum, but is noticeably different in the energy range $0.1\,\text{TeV} \lesssim E \lesssim 10\,\text{TeV}$, both in terms of spectral shape and intensity.    

Figure \ref{fig:t_90} shows the same spectra as those shown in Figure \ref{fig:t_45}, but at $\theta=90^{\circ}$, i.e., the plane where the neutral sheet is located.  Compared to the no-drift scenario, drift in the $qA>0$ scenario leads to a softer spectrum at $E\gtrsim 1\,\text{TeV}$, but a markedly harder spectrum at smaller energy values.  At these lower energies it is possible to fit the spectrum with the power-law $N\propto E^{-1.4}$.  For the $qA<0$ scenario the spectrum is again similar to the no-drift scenario when $E\gtrsim 10\,\text{TeV}$.  At lower energies a feature that resembles the inverse of the one seen in the $qA>0$ spectrum develops.  This feature is an enhancement of the one also visible in Figure \ref{fig:t_45}.  

After an intensive investigation it was found that gradient and curvature drift have a minimal effect on the evolution of the spectrum.  However, the large deviations from the no-drift spectrum in Figure \ref{fig:t_90}, for both the $q>0$ and $qA>0$ scenarios, were found to be related to neutral sheet drift.

\section{Summary}

In this paper it has been shown that diffusion can be used to explain the spatial evolution of the X-ray spectra observed from PWNe.  As the presented simulations were done for a steady-state system, it may be argued that these solutions are limited.  However, \cite{Vorster2013} has shown that a time-dependent solution of (\ref{eq:te}) does not lead to new spectral features.  Note that these time-dependent solutions also take into account the fact that the outer boundary of the PWN expands with time. 

While the simulations are presented within the context of PWNe, the results are also valid for any central source system.  This further implies that the evolution is not limited to electron/positron spectra, but can also be applied to protons, or even partially ionised nuclei. 

Lastly, it was found that gradient and curvature drift resulting from the Archimedean spiral geometry of the magnetic field have a negligible effect on the spectral evolution.  However, the presence of a neutral sheet, and as a result drift associated with this structure, can lead to significant modification of both the particle intensity and shape of the spectra.  For the $qA>0$ scenario neutral sheet drift leads to a hard spectrum at lower energies, possibly explaining the origin of the spectrum required to produce the hard radio synchrotron emission typically observed from PWNe (e.g., \cite{Weiler1980}).  This result might also be supported by the radio measurements of \cite{Dodson2003}.  Observing the inner region of Vela X, \cite{Dodson2003} found radio lobes in the same plane where the neutral sheet should also be located.  Furthermore, \cite{Dodson2003} derived a spectral index similar to the one predicted by the model.  It is therefore tentatively suggested that the radio lobes could be related to the presence of a neutral sheet.

\end{document}